\documentclass[aps,twocolumn,amsmath,floatfix, showpacs]{revtex4-1}
\usepackage{amssymb}
\usepackage{graphicx}
\usepackage{array}
\usepackage{hhline}
\usepackage{longtable}
\usepackage{bm}
\usepackage{hyperref}

\begin{document}

\newcommand{\rsp}[1]{\hspace{-0.15em}#1\hspace{-0.15em}}

\title{Evaporation-driven ring and film deposition from colloidal droplets}

\author{C. Nadir Kaplan$^1$ and L. Mahadevan$^{1, 2, 3, 4}$}
\affiliation{\\$^{1}$School of Engineering and Applied Sciences, Harvard University, Cambridge, MA 02138, USA.
\\$^{2}$Wyss Institute for Biologically Inspired Engineering, Harvard University, Boston, MA 02115, USA.
\\$^{3}$Kavli Institute for Bionano Science and Technology, Harvard University, Cambridge, MA 02138, USA.
\\$^{4}$Department of Physics, Harvard University, Cambridge, MA 02138, USA.}

\date{\today}

\begin{abstract}
Evaporating suspensions of colloidal particles lead to the formation of a variety of patterns, ranging from a left-over ring of a dried coffee drop to uniformly distributed solid pigments left behind wet paint. To characterize the transition between single rings, multiple concentric rings, broad bands, and uniform deposits, we investigate the dynamics of a drying droplet via a multiphase model of colloidal particles in a solvent. Our theory couples the inhomogeneous evaporation at the evolving droplet interface to the dynamics inside the drop, i.e. the liquid flow, local variations of the particle concentration, and the propagation of the deposition front where the solute forms an incompressible porous medium at high concentrations. A dimensionless parameter combining the capillary number and the droplet aspect ratio captures the formation conditions of different pattern types.
\end{abstract}
\maketitle

When coffee drops, soup splatter, salted snowmelt or other suspension dries out, the suspended solid remains as a residual stain or pattern. This is the result of a singular evaporative flux at the contact line which is pinned at the substrate \cite{Deegan1}. The resulting fluid flow advects particles to the edge where they aggregate, while the fluid itself evaporates. Consequently, single~\cite{Deegan1, Deegan2, Deegan3, Popov, Snoeijer1} or multiple rings~\cite{Kaplan, Adachi, Stone2, Stone3, Chang, Zhang, Kaya} form in the vicinity of the contact line.  In contrast, eliminating the surface roughness of the substrate~\cite{Snoeijer2}, or drying a suspension of anisotropic colloidal particles~\cite{Yodh1},  leads to the uniform deposition of particles over the droplet area. Alternatively, the transition from narrow single rings to uniform films can be engineered when the evaporation happens relatively fast~\cite{Klabunde, Narayanan, Bigioni}.

\begin{figure}
\centering
\includegraphics[width=1\columnwidth]{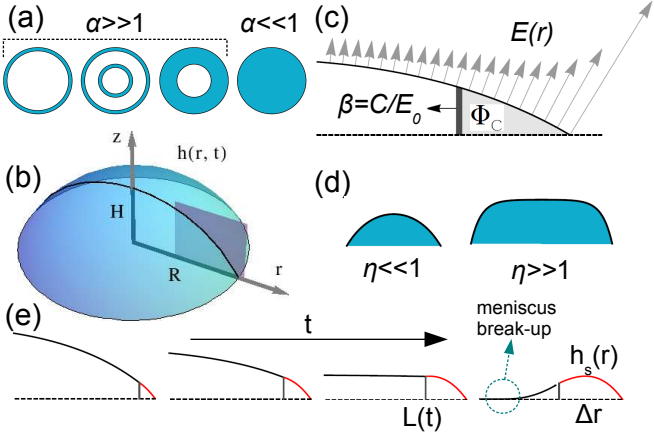}
\caption{Schematics of a drying suspension on a horizontal substrate. (a) Single rings, multiple concentric rings, broad bands, and uniform films are displayed. (b) The fluid height $h(r, t)$ and the polar coordinate system. The initial aspect ratio is $\epsilon\equiv H/R\,.$ (c) The droplet cross section (dark plane in (b)) and variable definitions. (d) As a function of $\eta\equiv R/\ell_{cap}$ ($l_{cap}\equiv \sqrt{\gamma/\rho g}$ : capillary length), the initial height profile $h(r, 0)$ of a small droplet ($\eta\ll 1$) when the Laplace pressure $-\gamma\kappa$ dominates the pressure $p$ and a puddle ($\eta \gg 1$) when the hydrostatic pressure $\rho g h$ is dominant. (e) The evolution of the deposit width $\Delta r$, the solid height $h_s(r)$ (red), and the interface position $L(t)$ in time $t\,.$	}
\label{fig:schematics}
\end{figure}

To understand the transition between single rings, multiple concentric rings, broad bands, and uniform deposits (Fig.~\ref{fig:schematics}(a)),  when the contact line of the drying drop (solvent viscosity $\mu$, interfacial tension $\gamma$) is pinned to the substrate, we note that in many situations the droplet aspect ratio $\epsilon\equiv H/R \ll 1$  {(Fig.~\ref{fig:schematics}(b))}. In this limit, the Navier-Stokes equations for fluid flow simplify via the lubrication approximation ~\cite{Oron}.  At a scaling level,  the pressure $p\sim \gamma H/R^2$ is determined by Laplace's law,  and  the viscous capillary flow speed of the liquid in the radial direction is $v_f\equiv\alpha E_0/\epsilon\,$, with $E_0$ being the evaporation rate, and $\alpha$ a dimensionless constant. Then, a balance between the radial pressure gradient $p/R$ and transverse viscous shear $\mu v_f/H^2$ implies that $\gamma H/R^3=\mu \alpha E_0/\epsilon H^2\,$. Thus, $\alpha\equiv \epsilon^4/Ca\,,$ where $Ca\equiv\mu E_0/\gamma$ is the capillary number, which characterizes the ratio of viscous to capillary forces. When $\alpha\gg 1\,,$ $v_f\gg E_0$, and the colloids are carried towards the contact line before the drying is complete, so that single ring, concentric rings, or a broad band forms, depending on the initial colloidal volume fraction $\Phi_0$. When $\alpha\ll 1\,,$ $v_f\ll E_0\,,$ and uniform films form as the solvent rapidly dries before most colloids can reach the droplet edge~\cite{Bigioni}. 

During deposition, colloidal particles arriving at the deposition front slow down and stop, while fluid continues to flow through the porous deposit. Hydrodynamically, in the interior of the droplet, the particle concentration is low, and pressure gradients are balanced by viscous stresses in Stokes flow~\cite{Landau} , while in the porous deposit, where the fluid concentration is low,  pressure gradients lead to flow through pores governed by Darcy's law~\cite{Landau}. Thus, we must account for a ``Stokes-Darcy transition" to characterize flow in the slender drying droplet as particles are carried by the fluid, before being eventually arrested, while the fluid evaporates away. The form of the residual patterns requires that we also consider the speed of the deposition front $C=C(\Phi_0, t)$ relative to the evaporation rate $E_0\,.$ Indeed, the type of patterning is governed by the dimensionless speed $\beta=\beta(t)\equiv C/E_0$ at the interface separating the liquid and forming deposit (Fig.~\ref{fig:
schematics}(c)). A uniform film forms when $\beta$ is sufficiently large to prevent the liquid meniscus, pinned to the elevated deposit edge, from touching down the substrate. Conversely, lower $\beta$ result{s} in single rings, multiple rings, or broad bands, following meniscus break-up (Fig.~\ref{fig:schematics}(e)). The coupling of evaporation-driven flow with a transition from a dilute suspension to porous plug requires a multiphase description of the process. 

In the lubrication approximation, the depth-averaged solute and solvent velocities are given by $V_{s}\equiv h^{-1}\int^h_0 v_{s}(r, z, t) dz$ and $V_{f}\equiv h^{-1}\int^h_0 v_{f}(r, z, t) dz$, where $v_{s}(r, z, t)$ and $v_{f}(r, z, t)$ are the local solute and solvent velocities, respectively. The depth-averaged solute volume fraction is $\Phi(r, t)=h^{-1}\int^h_0 \phi(r,z,t) dz$, where $\phi(r,z,t)$ is the local particle volume fraction, $1-\Phi(r, t)$ is the  {depth-averaged} solvent volume fraction and $h(r, t)$ is the droplet height (Fig.~\ref{fig:schematics}(b)). The growing deposit near the contact line due to the particle accumulation forms a porous plug with a volume fraction $\Phi_c$ of the particle close packing ($\Phi_c\approx 0.74$ for hexagonal packing in three dimensions). To characterize the transition from  the Stokes regime for dilute suspensions ($\Phi\ll 1$) to the Darcy regime in the porous medium ($\Phi\simeq\Phi_c$) we need a model that naturally transitions from one regime to another. A natural candidate is an interpolation between these two linear flow regimes via the Darcy-Brinkman equation~\cite{Brinkman}, which in the lubrication limit reads as
\begin{equation}
\frac{\partial p}{\partial r}=\mu \frac{\partial^2 v_f}{\partial z^2}-\frac{1}{k}\left(v_f-v_s\right)\,,
\label{eq:DarcyBrinkman}
\end{equation}
where $k$ is the permeability of the porous plug, and the pressure is given by $p=-2\gamma \kappa + \rho g h$ (Fig.~\ref{fig:schematics}(d)), with $\rho$ the suspension density, and $g$ being gravity.  {B}y defining $\tan{\theta}\equiv-\partial_r h$, the mean curvature $\kappa$ of the liquid-air interface is given by $\kappa=-\partial_r\left(r \sin{\theta}\right)/2r$. When $\Phi \ll 1$, we would like to ensure that the solute and solvent velocities coincide, i.e. $v_s \approx v_f$ as the particles are advected by the fluid in an approximately Stokesian regime. Beyond the deposition front, $v_s \rightarrow 0$ as $\Phi \rightarrow \Phi_c$, we would like to ensure that we recover the Darcy regime where the particles are arrested in the porous plug and the fluid velocity is proportional  {only} to the pressure gradient {~\cite{supplemental}}. A simple closure of Eq.~\eqref{eq:DarcyBrinkman} consistent with these limits is ~\cite{Cohen} $v_s=\left[1-\left(\Phi/\Phi_c\right)^\Gamma\right]v_f\,$, where the 
exponent $\Gamma$ determines the rapidity of the crossover between the two regimes. Eq.~\eqref{eq:DarcyBrinkman} in combination with our closure relation, subject to the stress-free and no-slip boundary conditions $\partial v/\partial z|_{z=h}=0$ and  $v(z=0, t)=0\,,$ yields the depth-averaged velocities
\begin{equation}
\label{eq:theory3}
V_f=\frac{1}{a^3\mu h} \frac{\partial p}{\partial r} \left(\tanh{ah}-a h\right)\,,\quad V_s=\left(1-a^2\mu k\right) V_f\,,
\end{equation}
where $a^2\equiv(\mu k)^{-1}(\Phi/\Phi_c)^\Gamma\,$, with $1/a$ being the effective pore size. 

We define a set of dimensionless variables as follows: horizontal coordinate  $r\equiv R\tilde{r}$, vertical height $h\equiv H\tilde{h}$, time $t \equiv H/E_0 \tilde{t}$, velocities $V_{s, f}\equiv (\epsilon^3\gamma/\mu \nu^3) \tilde{V}_{s, f}$, dimensionless pressure $p\equiv \left(\gamma\epsilon/R\right) \tilde{p}$, where $\tilde{p}\equiv-2\kappa+\eta h$, $\eta\equiv R/\ell_{cap}\,,$   $\ell_{cap}\equiv\sqrt{\gamma/\rho g}$ is the capillary length, and the dimensionless evaporation rate $\tilde{E}(r)\equiv E(r)/E_0$ (Fig.~\ref{fig:schematics}(b)). Furthermore, we let $\nu\equiv H/\sqrt{k\mu}\,,$ be a scaled initial height and the Peclet number $Pe\equiv E_0 R/D_s\,,$ with $D_s$ being the solute diffusivity. Dropping the tildes from the dimensionless quantities, and using an axisymmetric polar coordinate system in the rest frame, the depth-averaged  scaled equations of local fluid and solute mass conservation are given by
\begin{equation}
\label{eq:theory4}
\frac{\partial}{\partial t} \left[(1-\Phi) h\right]+\frac{1}{r}\frac{\partial \left(r Q_f\right)}{\partial r}=-E(r)\sqrt{1+(\epsilon \partial_r h)^2}\,, 
\end{equation}
\begin{equation}
\label{eq:theory5}
\frac{\partial}{\partial t} \left[\Phi h\right]+\frac{1}{r}\frac{\partial \left(r Q_s\right)}{\partial r}=0\,,
\end{equation}
where the scaled liquid flux $Q_f(r,t)=\frac{\alpha}{\nu^3}(1-\Phi) h V_f$  {($Q_f(r,t)=(1-\Phi)hV_f$ in dimensional units),} \  the particle flux $Q_s(r,t)=\frac{\alpha}{\nu^3}\Phi h V_s-Pe^{-1}\epsilon h \frac{\partial\Phi}{\partial r}$  {($Q_s(r,t)=\Phi h V_s-D_s h \frac{\partial\Phi}{\partial r}$ in dimensional units)}, and $E(r)\equiv 1/\sqrt{1-r}$ is the scaled singular form of the local evaporation rate (Fig.~\ref{fig:schematics}(c)) along the droplet surface~\cite{Deegan1, Deegan2}. The coupled sixth order system of Eqs.~\eqref{eq:theory3}--\eqref{eq:theory5} becomes a boundary value problem for $h(r, t)$ and $\Phi(r, t)\,,$ once appropriate boundary and initial conditions are given.

\begin{figure}
\centering
\includegraphics[width=1.05\columnwidth]{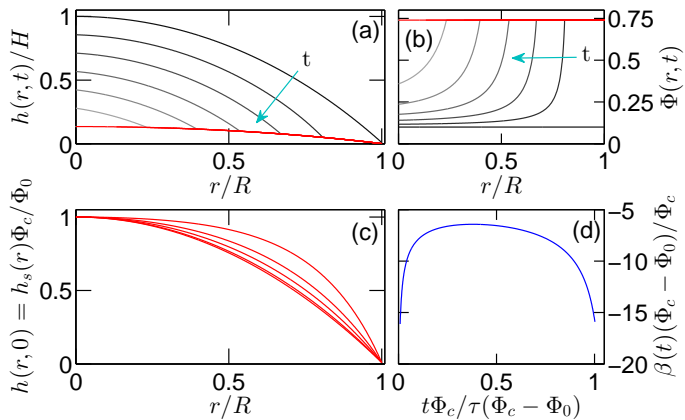}
\caption{Uniform film deposition ($\alpha\ll 1\,,$ $Pe\gg 1$). (a) The fluid interface $h(r, t)$ (grayscale) and the solid height $h_s(r)$ (red) evolution. (b) The change of the depth-averaged colloidal concentration $\Phi(r, t)$ (grayscale) corresponding to (a). The red line is at $\Phi_c=0.74\,.$ In (a) and (b), $\Phi_0=0.1\,,$ $\eta\equiv R/\ell_{cap}=1$, and the turquoise arrows denote the arrow of time with the grayscale changing from dark to light. (c) The relation between $h_s(r)$ and the initial hydrostatic profiles $h(r, 0)$ as a function of increasing $\eta\,,$ from bottom to top. (a), (b), and (c) are the solutions to Eqs.~\eqref{eq:theoryuniform1} and~\eqref{eq:theoryuniform2}. (d) The front propagation speed $\beta(t)=1/\epsilon \partial_r t_c(r)$ collapses on a single curve with the corresponding scaling of the axes.}
\label{fig:uniformfilm}
\end{figure}

\begin{figure*}
\centering
\includegraphics[width=1.0\textwidth]{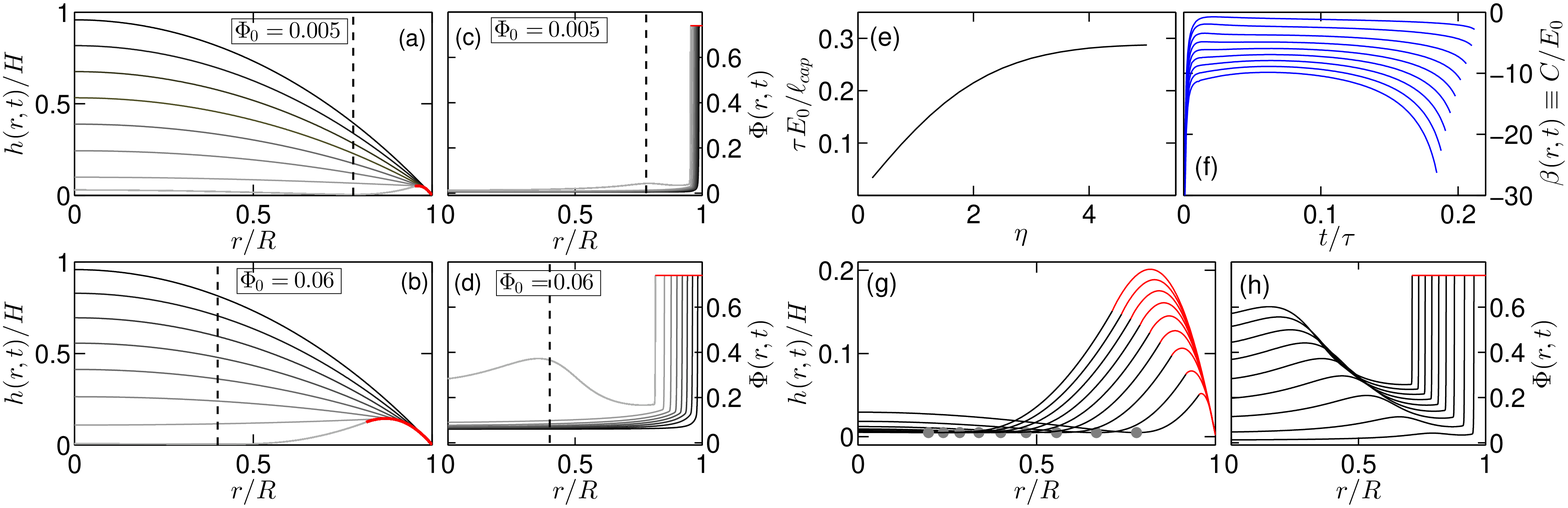}
\caption{Deposition of single rings and multiple concentric rings ($\alpha\gg 1$). (a), (b) The evolution of the interface height $h(r, t)$ (grayscale) and the solid height $h_s(r)$ (red) for the given initial particle volume fractions $\Phi_0$. (c), (d) The evolution of the local colloidal volume fraction $\Phi(r, t)$ corresponding to (a) and (b). The grayscale changes from dark to light with the increasing time. The red lines in (c) and (d) are $\Phi_c=0.74\,.$ The dashed lines denote the touch-down location $r_0$ of the meniscus. (e) The time scale $\tau$ of evaporation as a function of $\eta$ (see text). (f) The deposition speed $\beta(t)$ as a function of increasing $\Phi_0$ from top to bottom. (g) $h(r, t)$ (black), $h_s(r)$ (red), and (h) $\Phi(r, t)$ at the end of the droplet lifetime are shown as a function of increasing $\Phi_0$ from bottom to top. In (g), from right to left, $r_0$ (gray dots) decreases when $\Phi_0$ increases. The data in (f), (g), and (h) correspond to $\Phi_0\in\{5\times10^{-3}\cup\left[1.5\times10^{-2}\,, 0.12\right]\}$ in $1.5\times10^{-2}$ increments. In (a)--(d) and (f)--(h), $\eta=1\,.$ These results are solutions to Eqs.~\eqref{eq:theory3}--\eqref{eq:theory5}, subject to the boundary conditions given in  Eqs.~\eqref{eq:theoryBC2}--\eqref{eq:theoryBC4} at $r=L(t)$, and $Q_s=Q_f=\partial h/\partial r=0$ at $r=0\,.$ The physical parameters are given in the main text.}
\label{fig:multiphase1}
\end{figure*}

When $\alpha\ll 1\,,$  the initial droplet aspect ratio is small ($\epsilon\ll 1$). Furthermore, if particle diffusion is negligible ($Pe\gg 1$),  advection dominates diffusion so that a sharp deposition front forms, and the liquid and solid fluxes $Q_{f, s}$ in Eqs.~\eqref{eq:theory4} and~\eqref{eq:theory5} vanish. $Q_f$ vanishes at the droplet edge as well, even though the fluid velocity $V_f$ diverges near the contact line, since $h V_f$ stays constant there~\cite{Deegan1, Deegan2, Deegan3}. Then, Eqs.~\eqref{eq:theory3}--\eqref{eq:theory5} reduce to the initial value problem $\partial h/\partial t=-E(r)$ and $\partial\left[\Phi h\right]/\partial t=0\,,$ which yield
\begin{equation}
\label{eq:theoryuniform1}
h(r, t)=h(r, 0)-E(r)t\,,\quad\Phi(r, t)=\Phi_0 \frac{h(r, 0)}{h(r, t)}\,,
\end{equation}
where the initial condition $h(r, 0)$ is the hydrostatic height profile given the contact angle $\theta_{e, 0}$ at the droplet edge. Locally, $h(r, t)$ evolves only until the critical time $t_c=t_c(r)$ (Fig.~\ref{fig:uniformfilm}(a)), when the colloids get jammed at the deposition front $r=L(t_c)$ (see Fig.~\ref{fig:schematics}(e)) with $\Phi(r, t_c)=\Phi_c=0.74$ (Fig.~\ref{fig:uniformfilm}(b)). The deposit height is then fixed at $h_s(r)=h(r, 0)-E(r) t_c(r)$ for $t\geq t_c$ (Fig.~\ref{fig:uniformfilm}(a)). The critical time for jamming $t_c$ is determined from Eq.~\eqref{eq:theoryuniform1} when $\Phi(r, t_c)=\Phi_c\,.$ Then $h_s(r)$ and $t_c$ are given by
\begin{equation}
\label{eq:theoryuniform2}
h_s(r)=\frac{\Phi_0}{\Phi_c} h(r, 0)\,,\quad t_c(r)=\frac{\left(\Phi_c-\Phi_0\right)}{\Phi_c}\frac{h(r, 0)}{E(r)}\,.
\end{equation}
In Fig.~\ref{fig:uniformfilm}(c), we show  the relation between $h(r, 0)$ and $h_s(r)$  as a function of $\eta\equiv R/l_{cap}\,$; both $h_s(r)$ and $h(r, 0)$ become more convex close to the contact line as $\eta$ increases, corresponding to an approximate hydrostatically dominated pressure $p$ (see Fig.~\ref{fig:schematics}(d)). 

Having calculated the scaled droplet height $h(r, t)$, the solid volume fraction $\Phi(r, t)\,,$ and the deposit height $h_s(r)\,,$ we turn to the dynamics of the deposition front location $L(t)\,$ given by the condition $dr/dt|_{r=L(t)} = C$. Since all lengths are scaled by the thickness of the droplet $H$, and velocities are scaled by the evaporation rate $E_0$, in scaled form, the deposition front follows the relation
\begin{equation}
\label{eq:beta}
\frac{dr}{dt}\bigg|_{r=L(t)}=\epsilon \beta\,. 
\end{equation}
{At} $t=t_c(r)\,,$ {taking the reciprocal of both sides of Eq.~\eqref{eq:beta} yields} $\beta(t)=1/\epsilon \partial_r t_c(r)\,.$ $\beta$ diverges both initially when $h(r, t)\ll 1$ close to $r=R$, and at the end of drying when $\Phi(r, t)\sim\Phi_0$ close to $r=0$ (Fig.~\ref{fig:uniformfilm}(d)). Thus far we have assumed that solute diffusion is not important, i.e. $Pe \gg 1$, but when this is not the case, we have to consider the complete system of equations.


When $\alpha\gg 1\,,$  we need to specify seven boundary conditions over the interval $r\in\left[0, L(t)\right]$ to determine $h(r,t), \Phi(r,t)$ and $\beta(t)\,.$ Three boundary conditions at $r=0$ are given by symmetry: the vanishing particle flux $Q_s(0, t)=0\,,$ the vanishing liquid flux $Q_f(0, t)=0\,,$ and the flat meniscus height profile $\partial h/\partial r|_{r=0}=0\,.$ At the deposition front $r=L(t)\,,$  the liquid flux into the deposit at the interface must compensate for the loss of solvent via the evaporation over the deposit height $h_s(r)\,.$ The differential form of this condition in the frame co-moving with the wall at a speed $\beta$ becomes
\begin{equation}
\label{eq:theoryBC2}
\frac{1}{\epsilon\beta}\frac{\partial}{\partial t}\left[(1-\Phi)h\alpha V_f\right]=-E(r)\sqrt{1+\left(\epsilon\partial_r h\right)^2}\,,
\end{equation}
where Eq.~\eqref{eq:beta} is used to express the spatial derivative in terms of the time derivative. As the colloids are arrested, the particle flux inside the solid vanishes. Therefore, in the moving frame, the particle flux continuity at $r=L(t)$ is given by $Q_s-\epsilon\Phi h\beta=-\epsilon \Phi_c h\beta\,,$ which yields
\begin{equation}
\label{eq:theoryBC1}
\beta(t)=\frac{1}{\epsilon (\Phi-\Phi_c)}\left(\frac{\alpha}{\nu^3}\Phi V_s -Pe^{-1}\epsilon\frac{\partial \Phi}{\partial r}\right)\,.
\end{equation}
The height $h_i$ at the wall between the liquid and the incompressible deposit satisfies $\partial h_i/\partial t=0$ in the rest frame. In the moving frame, this condition translates to
\begin{equation}
\label{eq:theoryBC3}
\frac{\partial h_i(t)}{\partial t}=\frac{\epsilon \beta}{r}\frac{\partial \left(r h\right)}{\partial r}\bigg|_{r=L(t)}\,.
\end{equation}
We fix the particle volume fraction $\Phi_i$ at the interface as
\begin{equation}
\label{eq:theoryBC4}
\Phi_i\equiv\Phi_c-10^{-3}\,,
\end{equation}
to extract $\beta(t)$ asymptotically from Eq.~\eqref{eq:theoryBC1}.  Finally, the initial aspect ratio $\epsilon=\epsilon(\eta, \theta_{e, 0})$ is specified by the hydrostatic height $h(r, 0)$ (Fig.~\ref{fig:uniformfilm}(c)). The initial particle distribution is given by $\Phi(r, 0)=(\Phi_i-\Phi_0)e^{[r-L(0)]/d_0}+\Phi_0$, where $d_0\ll1$ and $L(0)=1\,.$

Using the physical parameters $\gamma\sim0.1$N/m, $\rho\sim10^3$kg/m$^3$, $g\sim 10$m/s$^2$, $E_0\sim10^{-6}$m/s, $\mu\sim 10^{-3}$Pa$\cdot$s, $D_s\sim 5\times 10^{-11}$m$^2$/s~\cite{Bodiguel} and $\theta_{e, 0}=15^{o}$, we find the capillary length $\ell_{cap}\sim10^{-3}$m and the time scale $\tau=\tau(\eta, \theta_{e, 0})\equiv H/E_0=\eta\epsilon(\eta, \theta_{e, 0})\ell_{cap}/E_0=\eta\epsilon 10^3$s (Fig.~\ref{fig:multiphase1}(e)). Then, $Ca\equiv \mu E_0/\gamma=10^{-8}$, and $Pe^{-1}\equiv D_s/E_0 R=5\times 10^{-2}$, so that $\alpha\gg 1\,.$  We choose $\nu=1$ since the effect of bigger $\nu$ on the formation of rings and bands is insignificant as long as $\alpha\gg \nu^2$~\cite{supplemental} holds. The exponent $\Gamma$ is chosen as $\Gamma=4\,,$ which models a narrow transition region between the two flow regimes as a function of $\Phi\,.$


We numerically solve Eqs.~\eqref{eq:theory3}--\eqref{eq:theory5} subject to the boundary conditions $Q_s=Q_f=\partial h/\partial r=0$ at $r=0$ and Eqs.~\eqref{eq:theoryBC2}--\eqref{eq:theoryBC4} at $r=L(t)$, for the droplet-air interface height $h(r, t)$, the depth-averaged particle volume fraction $\Phi(r, t)$, and the deposition front velocity $\beta(t)$ via the COMSOL finite element package~\cite{Comsol}. The divergence of the evaporation rate at the contact line is resolved by assuming $E(r)=1/\sqrt{1+\bar{\varepsilon}-r}$ when $r\rightarrow 1\,,$ where $\bar{\varepsilon}=0.01\,.$

In Figs.~\ref{fig:multiphase1}(a)--(b), the time evolution of $h(r, t)$ and $h_s(r)$ are shown in the rest frame. Towards the end of the droplet lifetime, the meniscus reverses curvature and touches the substrate at a location $r_0$ (dashed vertical lines in Figs.~\ref{fig:multiphase1} (a)--(d), and gray dots as a function of $\Phi_0$ in Fig.~\ref{fig:multiphase1} (g)), when the evaporation ceases at a time $t_f\sim 0.2\tau\,.$ The meniscus touch-down (see Fig.~\ref{fig:schematics}(e)), assumed to be axisymmetric~\cite{footnote1}, is followed by a break-up into two contact lines~\cite{Kaplan}. To regain the equilibrium contact angle $\theta_{e, 0}$~\cite{deGennes}, the contact line closer to $r=0$ moves to the droplet center, whereas the other one will move towards the deposition front and the fluid will be wicked by the jammed colloids.  

When $\Phi_0\ll 1\,,$ narrow rings form with a width $\Delta r(t)\equiv 1-L(t)\ll 1$ (Figs.~\ref{fig:multiphase1}(a), (c)).   The magnitude of the front speed $|\beta(t)|$, shown in Fig.~\ref{fig:multiphase1} (f), determines the width $\Delta r$~\cite{supplemental}. When $\Phi_0$ increases, $|\beta(t)|$ increases for all $t$ and diverges faster at $t_f\,.$ For single rings, $\beta(t)$ reaches a steady state at early times for $\Phi_0\ll1\,.$ For intermediate $\Phi_0\,,$ the particle concentration builds up below $r_0$, as evidenced by the maxima in $\Phi(r, t)$ at the liquid side of the wall in Figs.~\ref{fig:multiphase1}(d) and (h). When $\Phi_0$ increases, $r_0$ shifts to the droplet center, as demonstrated in Figs.~\ref{fig:multiphase1}(a)--(d), and (g)~\cite{supplemental}. Following meniscus break-up, the inner contact line moves until it is restrained by the meniscus~\cite{footnote2}, resulting in a new ring. In Fig.~\ref{fig:multiphase1}(b), we show the profiles shortly before the concentric inner ring formation, noting that $\beta$ changes rapidly over the course of drying for bigger $\Phi_0\,.$  When $\Phi_0$ is large, we see the formation of a broad band  -- Fig.~\ref{fig:multiphase1}(g), with $\Delta r\simeq R/3$ near the contact line, and the touch-down location $r_0\rightarrow 0$. 

The shape of the band $h_s(r)$ for all $\Phi_0$ is governed by Eq.~\eqref{eq:theoryBC3}: If the meniscus contact angle at the wall $\theta_e(L(t), t)>0\,,$ then $h_s(r)$ increases as the band expands (early stages), and $\theta_e(L(t), t)<0$ would lead to a decreasing height profile (late stages). The combination of the early- and late-stage behaviors result in the curved $h_s(r)$ profiles demonstrated in Figs.~\ref{fig:multiphase1}(a), (b), and (g). The maximum solid thickness $h_{max}$ is then achieved at the transition between the two regimes (the maxima of the red curves in Fig.~\ref{fig:multiphase1}(g)), as $h(r, t)$ changes curvature due to the evaporation~\cite{supplemental}. 

Our multiphase model of an evaporating colloidal droplet describes the different deposition patterns and the transitions between them by accounting for how the droplet evaporates inhomogeneously even as the dilute suspension transitions into a porous plug in the neighborhood of the contact line.  We are able to characterize these phenomena in terms of two parameters, the initial concentration $\Phi_0$ and the scaled evaporation rate $\alpha\,.$ When $\alpha\gg 1\,,$ we get single rings, multiple rings, and broad bands as $\Phi_0$ increases, while when $\alpha\ll 1$ uniform films assemble over the entire droplet area, consistent with observations. What remains is to understand the deposit thickness and the particle order in it, and will require coupling our multiphase macroscopic theory to a microscopic theory for particulate ordering in dense suspensions. 

We thank A. Carlson for fruitful discussions and T. Salez for a critical reading of the manuscript. This research was supported by the Air Force Office of Scientific Research (AFOSR) under Award FA9550-09-1-0669-DOD35CAP and the Kavli Institute for Bionano Science and Technology at Harvard University.


\begin{thebibliography}{}
\bibitem{footnote1} We ignore local meniscus touch-down due to the irregularities of the substrate, particle adhesion on the surface, and the disjoining pressure of the liquid.
\bibitem{footnote2} The particle aggregate close to the inner contact line in motion may either adhere to the substrate or be dragged by the contact line until it becomes jammed.
\end{thebibliography}
\end{document}


\newcommand{\rsp}[1]{\hspace{-0.15em}#1\hspace{-0.15em}}

\title{Supplementary information for ``Evaporation-driven ring and film deposition from colloidal droplets''}

\author{C. Nadir Kaplan$^1$ and L. Mahadevan$^{1, 2, 3, 4}$}
\affiliation{
\\$^{1}$School of Engineering and Applied Sciences, Harvard University, Cambridge, MA 02138, USA.
\\$^{2}$Wyss Institute for Biologically Inspired Engineering, Harvard University, Boston, MA 02115, USA.
\\$^{3}$Kavli Institute for Bionano Science and Technology, Harvard University, Cambridge, MA 02138, USA.
\\$^{4}$Department of Physics, Harvard University, Cambridge, MA 02138, USA.}

\date{\today}

\maketitle

\begin{figure}
\centering
\includegraphics[width=1.0\columnwidth]{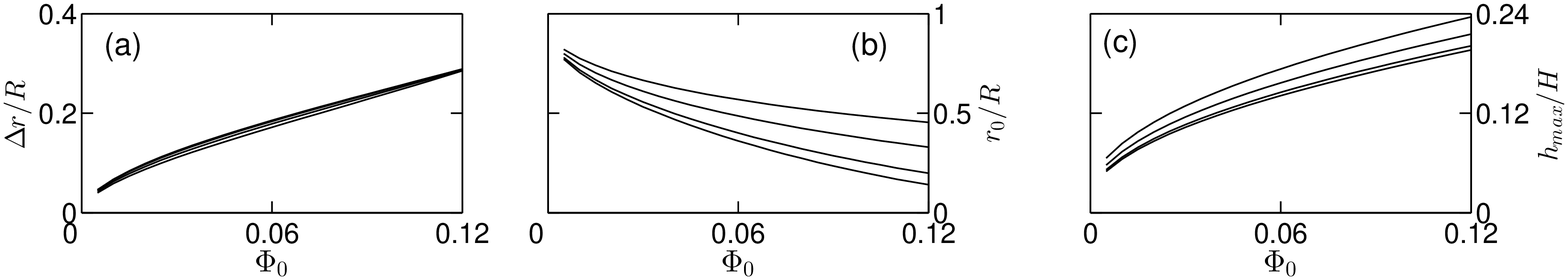}
\caption{Structural variables in the regime when $\alpha =\equiv \epsilon^4/Ca \gg 1$ regime, where $\epsilon=H/R$ is the aspect ratio of the drop, and $Ca= \mu E_0/\gamma$ is the Capillary number based on the evaporation rate $E_0$, the fluid viscosity $\mu$ and the surface tension $\gamma$.  (a)   the band width $\Delta r$, (b) the meniscus touch-down location $r_0\,,$ and (c) the maximum deposition height $h_{max}\,,$ as a function of $\Phi_0$ and $\eta\equiv R/\ell_{cap}\,.$ In (a) and (c) $\eta\equiv R/\ell_{cap}$ increases from bottom to top, where $\eta\in \{0.1, 1, 2, 3\}\,.$ When scaled over the droplet radius $R\,,$ $\Delta r$ collapses on a single curve.} 
\label{fig:multiphase2}
\end{figure}

\begin{figure}
\centering
\includegraphics[width=0.7\textwidth]{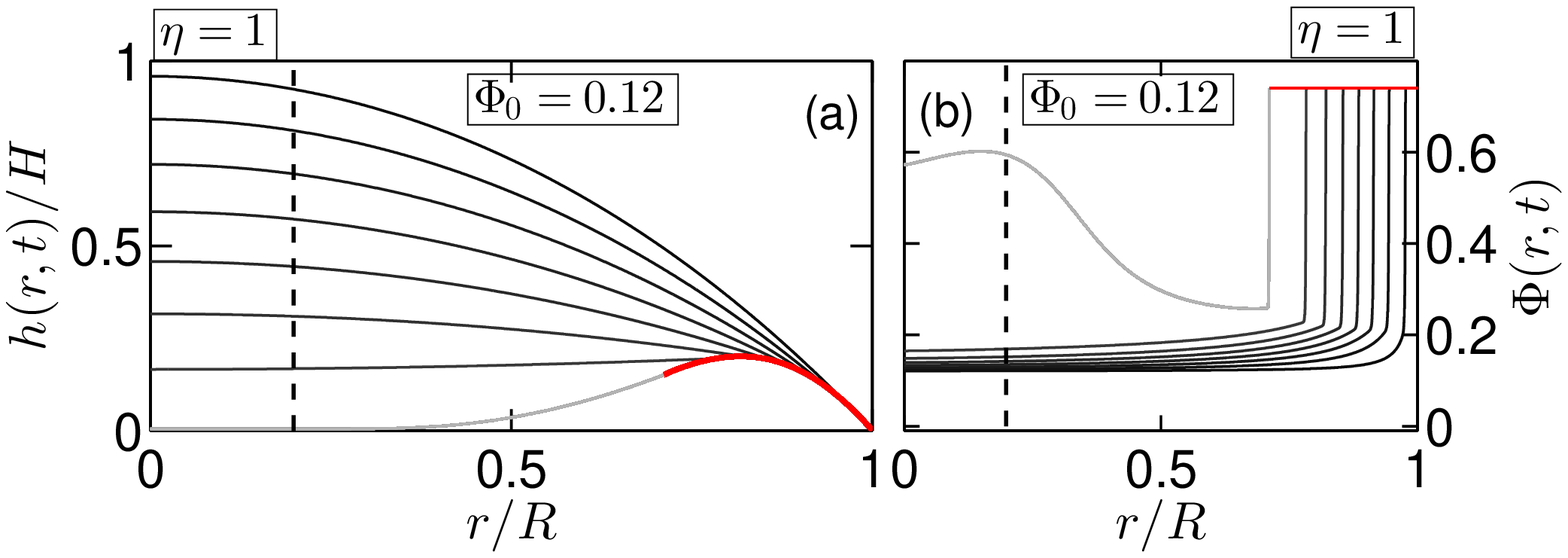}
\caption{Deposition of broad bands ($\alpha\gg 1$). (a) The interface height $h(r, t)$ evolution for the given initial particle volume fractions $\Phi_0$. (c) The evolution of the local colloidal volume fraction $\Phi(r, t)$ corresponding to (a) and (b). The grayscale changes from dark to light with the increasing time. The red curve in (a) is the forming solid-air interface $h_s(r)$. The red line in (c) is $\Phi_c=0.74\,.$ The dashed lines denote the touch-down location $r_0$ of the meniscus. These results are solutions to Eqs.~(2)--(4) in the main text, subject to the boundary conditions given in  Eqs.~(8)--(11) at $r=L(t)$ (see main text), and $Q_s=Q_f=\partial h/\partial r=0$ at $r=0\,,$ where $Q_s$ and $Q_f$ define the solid and liquid fluxes (Eqs.~\eqref{eq:theorySI1},~\eqref{eq:theorySI2} here). The physical parameters are given in the main text.}
\label{fig:multiphase3}
\end{figure}

\subsection{Effect of the dimensionless parameter $\nu$ on the solid and liquid fluxes}

The liquid flux $Q_f$ and the particle flux $Q_s$ in Eqs.~(3) and (4) in the main text are given by
\begin{equation}
\label{eq:theorySI1}
Q_f=Q_f(r, t)\equiv \frac{\alpha}{\nu^3}(1-\Phi) h V_f\,, 
\end{equation}
\begin{equation}
\label{eq:theorySI2}
Q_s=Q_s(r, t)\equiv \frac{\alpha}{\nu^3}\Phi h V_s-Pe^{-1}\epsilon h \frac{\partial\Phi}{\partial r}\,,
\end{equation}
where the dimensionless fluid and solid velocities read
\begin{equation}
\label{eq:theorySI3}
V_f=\frac{1}{h}\left(\frac{\Phi_c}{\Phi}\right)^{3 \Gamma/2} \frac{\partial p}{\partial r} \left\{\tanh{\left[\nu h \left(\frac{\Phi}{\Phi_c}\right)^{\Gamma/2}\right]}-\nu h \left(\frac{\Phi}{\Phi_c}\right)^{\Gamma/2} \right\}\,,\quad V_s=\left[1-\left(\frac{\Phi}{\Phi_c}\right)^{\Gamma}\right] V_f\,,
\end{equation}
As $\Phi\rightarrow 0$, the power series expansion of the dimensionless $V_{s, f}$ to the third order is given as
\begin{equation}
\label{eq:theorySI4}
V_f=\frac{1}{h}\left(\frac{\Phi_c}{\Phi}\right)^{3 \Gamma/2} \frac{\partial p}{\partial r} \left\{\nu h \left(\frac{\Phi}{\Phi_c}\right)^{\Gamma/2}-\frac{1}{3}\nu^3 h^3 \left(\frac{\Phi}{\Phi_c}\right)^{3\Gamma/2}-\nu h \left(\frac{\Phi}{\Phi_c}\right)^{\Gamma/2} \right\}\,,
\end{equation}
which, after simplifying, becomes
\begin{equation}
\label{eq:theorySI5}
V_f=-\nu^3\frac{h^2}{3} \frac{\partial p}{\partial r}\,.
\end{equation}
yielding the Stokes flow speed. Substituting Eq.~\eqref{eq:theorySI5} into Eq.~\eqref{eq:theorySI1}, the liquid flux $Q_f$ becomes
\begin{equation}
\label{eq:theorySI6}
Q_f\equiv \alpha(1-\Phi) h \left(-\frac{h^2}{3} \frac{\partial p}{\partial r}\right)\,, 
\end{equation}
Thus, the importance of the advective flux is solely measured by $\alpha$ in this limit. However, when the suspension becomes slurry-like at $\Phi_0\simeq\Phi_{\ast}$, and $\nu\gg 1$ corresponding to a small permeability $k$, the second term of $V_f$ in Eq.~\eqref{eq:theorySI3} becomes dominant, leading to
\begin{equation}
\label{eq:theorySI7}
V_f=-\nu\left(\frac{\Phi_c}{\Phi}\right)^{\Gamma} \frac{\partial p}{\partial r}\,,
\end{equation}
and the prefactor of $V_{s, f}$ in Eqs.~\eqref{eq:theorySI1} and~\eqref{eq:theorySI2} reduces to $\alpha/\nu^2\,.$ Then, for a slurry-like mixture with a small $k$, $\alpha \gg \nu^2$ leads to both single or concentric rings, as well as broad bands, whereas $\alpha \ll \nu^2$ results in uniform deposition. 

\subsection{Movies}
\begin{enumerate}
 \item Movie 1: The evolution of the liquid meniscus $h(r, t)\,,$ solid height $h_s(r)$ and the local volume fraction $\Phi(r, t)$ of an evaporating droplet where $\Phi_0=5\times10^{-3}$ and $\eta=2\,.$
 \item Movie 2: The evolution of the liquid meniscus $h(r, t)\,,$ solid height $h_s(r)$ and the local volume fraction $\Phi(r, t)$ of an evaporating droplet where $\Phi_0=0.05$ and $\eta=2\,.$
 \item Movie 3: The evolution of the liquid meniscus $h(r, t)\,,$ solid height $h_s(r)$ and the local volume fraction $\Phi(r, t)$ of an evaporating droplet where $\Phi_0=0.12$ and $\eta=2\,.$
\end{enumerate}